\documentstyle[PASJadd]{PASJ95}

\begin{document}

\title{Roles of SNIa and SNII in ICM Enrichment}

\author{Yuhri {\sc Ishimaru}\\
    {\it Department of Astronomy, School of Science, The University of
    Tokyo, Bunkyo-ku, Tokyo 113}\\
    {\it E-mail(YI): ishimaru@astron.s.u-tokyo.ac.jp}\\
    and\\
    Nobuo {\sc Arimoto}\\
    {\it Institute of Astronomy, Faculty of Science,
	University of Tokyo, 2-21-1 Osawa, Mitaka, Tokyo 181}\\
    {\it Institut f\"ur Theoretische Astrophysik, Universit\"at Heidelberg,
    Tiergartenstrasse 15, D-69121 Heidelberg, Germany}\\
    {\it and}\\
    {\it Department of Physics, University of Durham, South Road,
    Durham, DH1 3LE, U.K.}}

\abst{
Based on ASCA observations Mushotzky et al. (1996, ApJ 466, 686) have recently
derived the relative-abundance ratios of $\alpha$-elements to
iron, [$\alpha$/Fe]$\simeq$0.2--0.3, 
for four rich clusters, and have suggested
that the origin of metals in an intra-cluster medium (ICM) is 
not a type-Ia supernovae (SNIa), but a type-II supernovae (SNII). 
However, these authors used the
solar {\it photospheric} iron abundance for ASCA data reduction,
while the {\it meteoritic} 
iron abundance is usually adopted in 
chemical-evolution studies.
It is true that although the photospheric and meteoritic 
solar abundances are consistent for most of the elements,
a serious discrepancy is known to exist for iron; indeed,
the photospheric abundance of iron is 
$N_{\rm Fe}/N_{\rm H} = 4.68\cdot 10^{-5}$ by number,
while the meteoritic value is $3.24\cdot 10^{-5}$.
The argument concerning the relative roles of SNIa and SNII
in ICM enrichment 
is quite sensitive to the precise values of [$\alpha$/Fe], and 
one should use an identical solar iron abundance in data reduction
as well as in theoretical arguments.  We therefore adopt the
meteoritic iron abundance, which is consistent with chemical-evolution
studies, and shift
Mushotzky et al.'s ASCA data by $\Delta$[$\alpha$/Fe]$\simeq -0.16$ dex.
By comparing the corrected [$\alpha$/Fe] values with
theoretical nucleosynthesis prescriptions of SNIa and SNII,
we reach a conclusion that an SNIa iron contribution of 50\% or higher
in the ICM enrichment
could not be ruled out, and might indeed be favoured based on the ASCA
spectra. 
}

\kword{
Galaxies: elliptical --- 
Galaxies: evolution --- 
Galaxies: intergalactic medium --- 
Galaxies: X-rays ---
Supernovae
}

\maketitle
\thispagestyle{headings}

\section{Introduction}

Clusters of galaxies are surrounded by hot X-ray emitting ICM
enriched with a large amount of iron (Rothenflug, Arnaud 1985;
Hatsukade 1989; Edge, Stewart 1991; Arnaud et al. 1992; Ikebe et 
al. 1992; Tsuru 1993). The universal ratio
of the iron mass in the ICM to the total galaxy 
luminosity (i.e., so-called IMLR; Arnaud et al. 
1992; Renzini et al. 1993) suggests that most of 
the iron was synthesized in cluster ellipticals, and ejected {\it via} 
SN-driven winds (e.g., Larson 1974; Arimoto, Yoshii 1987).

The relative-abundance ratios of $\alpha$-elements to iron in the ICM is 
one of the most important observational constraints on 
chemical-evolution models, because it directly tells the relative roles 
of SNIa and SNII in the enrichment of ICM. Recently, based on 
the ASCA X-ray spectra of four rich clusters of galaxies, Abell 496,
Abell 1060, Abell 2199, and AWM7, 
Mushotzky et al. (1996) reported on a rather unexpected result, which shows that 
[$\alpha$/Fe]$\simeq$ 0.2--0.3 in the ICM of these clusters.
The authors claim that these [$\alpha$/Fe] ratios are consistent
with the origin of {\it all} metals in SNII. 

The observed trend of [$\alpha$/Fe] vs [Fe/H] of G-dwarfs in the 
solar-neighbourhood of our Galaxy can be explained based on the 
different nucleosynthesis yields and lifetimes of SNIa and 
SNII (cf. Greggio, Renzini 1983; Wheeler et al. 1989; 
Matteucci, Fran\c cois 1992; Edvardsson et al. 1993).
A detailed model suggests
that 57\% of iron in the Sun should come from SNIa (Tsujimoto 
et al. 1995). The number ratio of SNIa progenitors to those of SNII 
is roughly determined by the initial-mass function (IMF) and a 
frequency of binaries in the mass interval corresponding to that of 
SNIa progenitors. If the IMF and binary frequency are the same 
in the solar neighbourhood and in elliptical galaxies, and if most 
of the interstellar medium of ellipticals is eventually mixed with the 
ICM, a significant amount of iron in the ICM must come from SNIa. 
Indeed, previous chemical-evolution studies 
suggested that the ICM was enriched mainly by SNIa (Matteucci,
Vettolani 1988; Renzini et al. 1993; Mihara, Takahara 1994).
Thus, if the interpretation of the ASCA data by Mushotzky et 
al. is correct, it would require a somewhat complicated mechanism 
for enriching the ICM and/or a completely different understanding
concerning SNIa nucleosynthesis.

Recent theoretical studies have attempted to interpret the observed 
{\it high} values of [$\alpha$/Fe]. \ Elbaz et al. (1995) suggest a bimodal 
star-formation model, in which the formation of SNIa progenitors is strongly 
suppressed, originally proposed by Arnaud et al. (1992).
Matteucci, Gibson (1995) find that the SNIa products 
should remain in the halo of ellipticals, since the thermal energy 
input from SNIa is not sufficient to induce a late-time wind. 

In this paper we argue that Mushotzky et al. (1996) 
might not interpret the
ASCA data properly, because the iron abundance in the solar photosphere
is used when the
authors estimate the abundances of the X-ray emitting gas by fitting a 
spectral synthesis model to the ASCA data. Although the 
solar abundances, estimated from either the solar photosphere or 
meteorites, are almost consistent for most of the elements, a serious 
discrepancy is known to exist concerning the iron abundance (Anders, Grevesse 1989).
On the other hand, 
studies of stellar nucleosynthesis use the solar abundances
derived from the meteorites by Anders and Grevesse (1989)
to explain the observed [$\alpha$/Fe] of low-metal G-dwarf stars
(e.g., Thielemann et al. 1993, 1996); also, galactic chemical-evolution
studies use the meteoritic abundances to explain the observed
trend of [$\alpha$/Fe] vs [Fe/H] in the solar neighbourhood
(e.g., Timmes et al. 1995; Tsujimoto et al. 1995).
Moreover, the meteoritic solar abundances are generally used
in chemical-evolution studies, including recent studies concerning
ICM enrichment (e.g., Elbaz et al. 1995; Matteucci, Gibson 1995).
As we shall demonstrate in the following sections, the iron-abundance
discrepancy between the solar photosphere and meteorites is not negligibly
small; thus, for studying the origin of iron in the ICM
one should use the same iron abundance in the data reduction
as well as in theoretical modeling.

We therefore modified Mushotzky et al.'s (1996) relative-abundance 
ratios with the meteoritic solar iron abundance (Anders, Grevesse 
1989). By using the modified data and the {\it stellar} yields of 
SNIa and SNII nucleosynthesis, we find that the observed 
relative abundances are consistent with an origin of 
iron more than half, perhaps even more, in SNIa and less in SNII.\  
We will present a more extended study concerning cluster chemical-evolution
in a companion paper.

In section 2 we discuss the solar iron-abundance discrepancy and
apply the meteoritic values to the relative-abundance ratios of 
the ICM of the four clusters of galaxies studied by Mushotzky et al. 
(1996). In section 3 we calculate the SNIa and SNII {\it stellar} 
yields in order to derive the fractional contribution of the SNIa products to the ICM.
\ Discussions and our conclusions are given in section 4, and section 5, 
respectively.

\section{Solar Iron Abundance and [$\alpha$/Fe] of the ICM}

\newcommand{\lw}[1]{\smash{\lower2.ex\hbox{#1}}}
\begin{table*}[t]
\small
\begin{center}
Table~1.\hspace{4pc}
 Abundances of ICM normalized by the meteoritic abundances.\\
\end{center}
\vspace{6pt}
\begin{tabular*}{\textwidth}{@{\hspace{\tabcolsep}
\extracolsep{\fill}}*{10}{l@{\hspace{\tabcolsep}}}}
\hline
\hline
 	 &\multicolumn{2}{c}{\lw{Abell 496}}   & \multicolumn{2}{c}{\lw{Abell 1060}} 
         & \multicolumn{2}{c}{\lw{Abell 2199}} &\multicolumn{2}{c}{\lw{AWM 7}}     & \lw{Average}\\
\multicolumn{10}{c}{ }\\
\hline
\multicolumn{10}{c}{\lw{SIS}}\\
\multicolumn{10}{c}{ }\\
\hline 
$[{\rm O/H}]$ & $-0.17$ &($-0.47, +0.05$)      & $-0.47$ &($-0.72, -0.31$)
             & $-0.36$ &($-0.74, -0.15$)      & $-0.32$ &($-0.68, -0.11$)      & $-0.32$\\
$[{\rm Ne/H}]$& $-0.04$ &($-0.22, +0.11$)      & $-0.24$ &($-0.39, -0.14$)
             & $-0.17$ &($-0.42, -0.01$)      & $-0.47$ &($-0.92, -0.24$)      & $-0.21$\\
$[{\rm Mg/H}]$& $-0.33$ &($-0.75, -0.11$)      & $-0.70$ &($-1.31, -0.45$)
             & $-0.34$ &($-0.77, -0.11$)      &  ---    &($     < -0.70$)      &  ---    \\
$[{\rm Si/H}]$& $-0.26$ &($-0.41, -0.13$)      & $-0.25$ &($-0.32, -0.18$)
             & $-0.04$ &($-0.14, +0.04$)      & $-0.28$ &($-0.39, -0.18$)      & $-0.19$\\
$[{\rm S/H}]$ & $-0.55$ &($-0.94, -0.34$)      & $-0.69$ &($-0.94, -0.54$)
             & $-0.61$ &($-1.15, -0.01$)      & $-0.78$ &($-1.28, -0.54$)      & $-0.66$\\
$[{\rm Ar/H}]$& $-0.50$ &($     < -0.12$)      & $-1.70$ &($     < -0.62$)
             &  ---    &($     < -0.61$)      &  ---    &($     < -0.83$)      &  ---    \\
$[{\rm Ca/H}]$& $-0.57$ &($     < -0.12$)      &  ---    &($     < -0.62$)
             &  ---    &($     < -0.39$)      &  ---    &($     < -0.70$)      &  ---    \\
$[{\rm Fe/H}]$& $-0.32$ &($-0.38, -0.24$)      & $-0.39$ &($-0.44, -0.26$)
             & $-0.30$ &($-0.35, -0.25$)      & $-0.31$ &($-0.36, -0.27$)      & $-0.33$\\
$[{\rm Ni/H}]$& $+0.17$ &($-0.07, +0.33$)      & $-0.16$ &($-0.72, +0.22$)
             & $+0.08$ &($-0.30, +0.25$)      & $-0.12$ &($-0.51, +0.09$)      & $+0.01$\\
\hline
$[{\rm O/Fe}]$& $+0.15$ &($-0.15, +0.36$)      & $-0.08$ &($-0.33, +0.08$)
             & $-0.06$ &($-0.45, +0.15$)      & $-0.01$ &($-0.37, +0.20$)      & $+0.01$\\
$[{\rm Ne/Fe}]$& $+0.28$&($+0.10, +0.43$)      & $+0.15$ &($+0.00, +0.25$)
             & $+0.12$ &($-0.12, +0.28$)      & $-0.16$ &($-0.61, +0.07$)      & $+0.12$\\
$[{\rm Mg/Fe}]$& $-0.01$&($-0.43, +0.21$)      & $-0.31$ &($-0.92, -0.06$)
             & $-0.05$ &($-0.48, +0.18$)      &  ---    &($     < -0.39$)      &  ---    \\
$[{\rm Si/Fe}]$& $+0.06$&($-0.09, +0.18$)      & $+0.14$ &($+0.07, +0.21$)
             & $+0.26$ &($+0.16, +0.33$)      & $+0.03$ &($-0.08, +0.13$)      & $+0.14$\\
$[{\rm S/Fe}]$& $-0.23$ &($-0.62, -0.02$)      & $-0.30$ &($-0.55, -0.15$)
             & $-0.31$ &($-0.86, +0.28$)      & $-0.47$ &($-0.97, -0.23$)      & $-0.32$\\
$[{\rm Ar/Fe}]$& $-0.18$&($     < +0.20$)      & $-1.31$ &($     < -0.23$)
             &  ---    &($     < -0.31$)      &  ---    &($     < -0.52$)      &  ---    \\
$[{\rm Ca/Fe}]$& $-0.25$&($     < +0.20$)      &  ---    &($     < -0.23$)
             &  ---    &($     < -0.09$)      &  ---    &($     < -0.39$)      &  ---    \\
$[{\rm Ni/Fe}]$& $+0.49$&($+0.25, +0.65$)      & $+0.23$ &($-0.33, +0.61$)
             & $+0.37$ &($+0.00, +0.54$)      & $+0.19$ &($-0.19, +0.40$)      & $+0.34$\\
\hline
\multicolumn{10}{c}{\lw{GIS}}\\
\multicolumn{10}{c}{ }\\
\hline
$[{\rm Si/H}]$& $-0.07$ &($-0.27, +0.07$)      & $-0.26$ &($-0.42, -0.12$)
             & $-0.004^\ast$&($-0.13, +0.12$) & $-0.12$ &($-0.50, +0.00$)      &   \\
$[{\rm S/H}]$ & $-0.44$ &($-1.28, -0.17$)      & $-0.61$ &($-1.15, -0.36$)
             & $-0.59$ &($     < -0.28$)      & $-0.28$ &($-0.52, -0.11$)      &   \\
$[{\rm Ar/H}]$&  ---    &($     < -0.13$)      & $-1.70$ &($     < -0.44$)
             &  ---    &($     < -0.15$)      &  ---    &($     < -0.41$)      &   \\
$[{\rm Ca/H}]$& $-0.02$ &($-0.90, +0.25$)      & $-0.64$ &($     < -0.15$)
             &  ---    &($     < -0.05$)      & $-0.49$ &($     < -0.03$)      &   \\
$[{\rm Fe/H}]$& $-0.31$ &($-0.38, -0.25$)      & $-0.29$ &($-0.36, -0.22$)
             & $-0.33$ &($-0.38, -0.27$)      & $-0.25$ &($-0.29, -0.19$)      &   \\
$[{\rm Ni/H}]$& $-0.66$ &($     < +0.13$)      & $-1.04$ &($     < -0.01$)
             & $+0.18$ &($-0.35, +0.42$)      & $+0.29$ &($-0.02, +0.48$)      &   \\
\hline
$[{\rm Si/Fe}]$& $+0.23$&($+0.03, +0.37$)      & $+0.04$ &($-0.13, +0.18$)
             &  $+0.33$&($+0.20, +0.45$)      & $+0.12$ &($-0.25, +0.24$)      &   \\
$[{\rm S/Fe}]$& $-0.14$ &($-0.97, +0.14$)      & $-0.31$ &($-0.86, -0.06$)
             & $-0.26$ &($     < +0.06$)      & $-0.03$ &($-0.28, +0.14$)      &   \\
$[{\rm Ar/Fe}]$&  ---   &($     < +0.17$)      & $-1.41$ &($     < -0.14$)
             &  ---    &($     < +0.18$)      &  ---    &($     < -0.17$)      &   \\
$[{\rm Ca/Fe}]$& $+0.28$&($-0.60, +0.55$)      & $-0.34$ &($     < +0.14$)
             &  ---    &($     < +0.28$)      & $-0.24$ &($     < +0.22$)      &   \\
$[{\rm Ni/Fe}]$& $-0.35$&($     < +0.44$)      & $-0.75$ &($     < +0.29$)
             & $+0.51$ &($-0.02, +0.75$)      & $+0.54$ &($+0.23, +0.72$)      &   \\
\hline
\end{tabular*}

\vspace{6pt}
$*$
{The original table of Mushotzky et al. seems to have an typographical
error in the GIS silicate abundance for Abell 2199. We therefore assume
[Si/H]$\simeq 0$ 
for this cluster (Y. Fukazawa, private communication).}
\end{table*}

The elemental abundances of 
astrophysical objects are usually expressed by the relative
values to the solar abundances. The so-called solar abundances  
have two alternative definitions: 
1) those abundances based on the meteorites and 
2) those based on the solar photospheric values.
The meteoritic abundances have converged
to the point where most elements are known to better than 10\%
(Anders, Grevesse 1989 and references therein).
The solar photospheric values have also been considerably improved;
Anders and Grevesse (1989) give quite
consistent values for most of the accurately determined elements 
with respect to the meteoritic abundances.
Nevertheless, significant discrepancies between the solar photosphere
and the meteorites still remain in several important elements, such as 
Fe, Mn, Ge, and Pb.\  In particular, the 
solar photospheric abundance of $^{26}$Fe is as high as 
$\log N_{\rm Fe}=7.67\pm0.03$ 
(where $\log N_{\rm H}=12$ and therefore $N_{\rm Fe}/N_{\rm H}=4.68~ 10^{-5}$)
by number, while a meteoritic analysis gives a much lower value of
$\log N_{\rm Fe}=7.51\pm0.01$ ($N_{\rm Fe}/N_{\rm H}=3.24~ 10^{-5}$)
(Anders, Grevesse 1989). This discrepancy had already 
been known previously: {\it high} iron abundances in the solar 
photosphere (e.g., Allen 1973;
Grevesse 1984a, b) and {\it low} values in the meteorites
(e.g., Cameron 1970, 1982; Anders, Ebihara 1982).

The iron abundance of the solar photosphere was recently re-determined
by Pauls et al. (1990), Holweger et al. (1990), and Bi\'emont et al. 
(1991) based on the Fe II lines, though it has often been determined
from the Fe I lines
in the past (e.g., Blackwell et al. 1984). 
Fe II ion is the dominant stage in the solar photosphere, and
its abundance is much less dependent on modeling and non-LTE effects (Bi\'emont
et al. 1991). Pauls et al. (1990) derived a {\it high} value of 
$\log N_{\rm Fe} = 7.66 \pm 0.06$ from a sample of three Fe II infrared lines,
in contrast with the {\it low} values of $\log N_{\rm Fe} = 7.48 \pm 0.09$ 
by Holweger et al. (1990) and $\log N_{\rm Fe} = 7.54 \pm 0.03$ 
by Bi\'emont et al. (1991), both from extensive samples of Fe II lines. 
Perhaps the {\it high} iron abundance found by Pauls et al. (1990) is not
statistically significant (see Holweger et al. 1990). Somewhat similar 
{\it low} result of $\log N_{\rm Fe} = 7.50 \pm 0.07$ has been derived by
Holweger et al. (1991), who re-determined the solar photospheric iron abundance
by using the Fe I lines along with newly calculated oscillator strengths. 
It thus seems
that the iron abundances of the solar photosphere and the meteorites 
(C1 chondrites) are identical within the uncertainties of the observations. We note
that this {\it low} iron abundance of the solar photosphere is much lower than
the solar iron abundance that Mushotzky et al. (1996) have adopted.

Because the {\it low} iron abundance is supported by 
these recent studies,
the meteoritic value of $\log N_{\rm Fe} \sim 7.50$ has been adopted in 
recent estimates of the iron abundance of G-dwarfs in the
solar neighbourhood (Edvardsson et al. 1993;
King 1993; Nissen et al. 1994). Norris et al. (1993) also mentioned that
their [Fe/H] values for field stars would increase 
when the {\it low} iron abundance of the Sun is proven to be correct.
Thus, in this paper we use the iron abundance derived from the 
meteorites, $\log N_{\rm Fe} = 7.51$, together with meteoritic abundances of
other elements,
and re-scale the ratios of the relative abundance of the elements of
the ICM for the clusters of galaxies reported by Mushotzky et al. (1996).
The resulting ratios are given in table 1. 
The error bars indicate 90\% confidence intervals.
As shown in table 1,
[Fe/H] should increase by $\sim 0.16$ dex and {[$\alpha$/Fe]} should decrease 
by the same amount.\ It is no longer evident whether the heavy elements 
in the ICM are all produced in SNII.\  In the next section we estimate 
quantitatively the fractional contribution from SNIa and SNII to the 
enrichment in clusters of galaxies 
by using the averaged {\it stellar} yields of SNIa and SNII.


\newpage
\section{SNIa vs SNII}

In section 2 we show that the relative abundances [$\alpha$/Fe]
of the ICM given by Mushotzky et al. (1996) should be decreased by 0.16 dex.
The resulting [$\alpha$/Fe] ratios are nearly {\it solar}, strongly suggesting
that the heavy elements in the ICM should be a mixture of SNIa and SNII ejecta.
We therefore consider the role of SNIa and SNII concerning
the nucleosynthesis in clusters of galaxies by using the latest 
stellar nucleosynthesis prescriptions.

We take SNII nucleosynthesis data from Thielemann et al. (1996)
for stars of 13, 15, 20, and 25 $M_\odot$, and from 
Tsujimoto et al. (1995) for 18, 40, and 70 $M_\odot$ stars.
For SNIa nucleosynthesis data, we use table 2 of Tsujimoto et al. 
(1995), who give the updated W7 model of Nomoto et al. (1984), 
calculated with the latest nuclear reaction rates by Thielemann et 
al. (1993).

The mass of the {\it i}-th element produced in a single event, i.e., 
the {\it stellar} yield (Tinsley 1980), of SNIa $y_{i,{\rm SNIa}}$ is 
constant irrespective of the details concerning progenitors, 
while that of SNII $y_{i,{\rm SNII}}(m)$ depends on the 
progenitor mass $m$. Thus, to estimate the relative contribution of SNIa 
and SNII to the enrichment of various elements in the ICM, we 
introduce the average {\it stellar} yield of SNII
taken for the mass range of SNII progenitors $m_{\rm l}$--$m_{\rm u}$:

\begin{equation}
\langle y_{i,{\rm SNII}}\rangle = 
  { \int^{m_{\rm u}}_{m_{\rm l}} y_{i,{\rm SNII}}(m)\phi(m) m^{-1}dm
   \over \int^{m_{\rm u}}_{m_{\rm l}}\phi(m) m^{-1}dm }.
\end{equation}

In equation (1), a Salpeter-like IMF, i.e., $\phi (m) \propto 
m^{-x}$ is assumed. We adopt the Salpeter IMF ($x=1.35$) and a flat 
IMF ($x=0.95$) from Arimoto and Yoshii (1987). In this paper the 
lower and upper mass limits of SNII progenitors $m_{\rm l}$, $m_{\rm u}$ are 
assumed to be $10M_\odot$ and $50M_\odot$, respectively.
Although the exact value of $m_{\rm l}$ is uncertain, because it strongly 
depends on the mass-loss rate during the AGB phase of 8--10$M_\odot$ 
stars, one can safely assume that SNII nucleosynthesis from 
low-mass stars of $m < 10M_{\odot}$ is negligible (Hashimoto et al. 
1993; Tsujimoto et al. 1995). 
We assume that the {\it stellar} yields decrease linearly in mass from 
$13M_{\odot}$ to $10M_{\odot}$. 
The upper mass limit $m_{\rm u}$ 
corresponds to a critical mass of stars which form black holes.
We adopt the value of $50 M_\odot$ derived by Tsujimoto et al. (1994).
The fractional contribution of SNIa to the ICM iron is then given as
\begin{equation}
{M_{\rm Fe,SNIa} \over M_{\rm Fe,total} }
   = {\zeta y_{\rm Fe,SNIa} 
      \over \zeta y_{\rm Fe,SNIa} + (1-\zeta) \langle y_{\rm Fe,SNII}\rangle},
\end{equation}
where $\zeta$ indicates the relative frequency of SNIa. With a help of
equation (2), we calculate the abundance ratios [Z$_i$/Fe] 
for Z$_i$ = O, Ne, Mg, Si, S, Ar, Ca, and Ni as a function of
$M_{\rm Fe,SNIa} / M_{\rm Fe,total}$. The solar abundances derived from 
the meteorites are taken from Anders and Grevesse (1989).

\begin{fv}{1}{27pc}%
\includegraphics{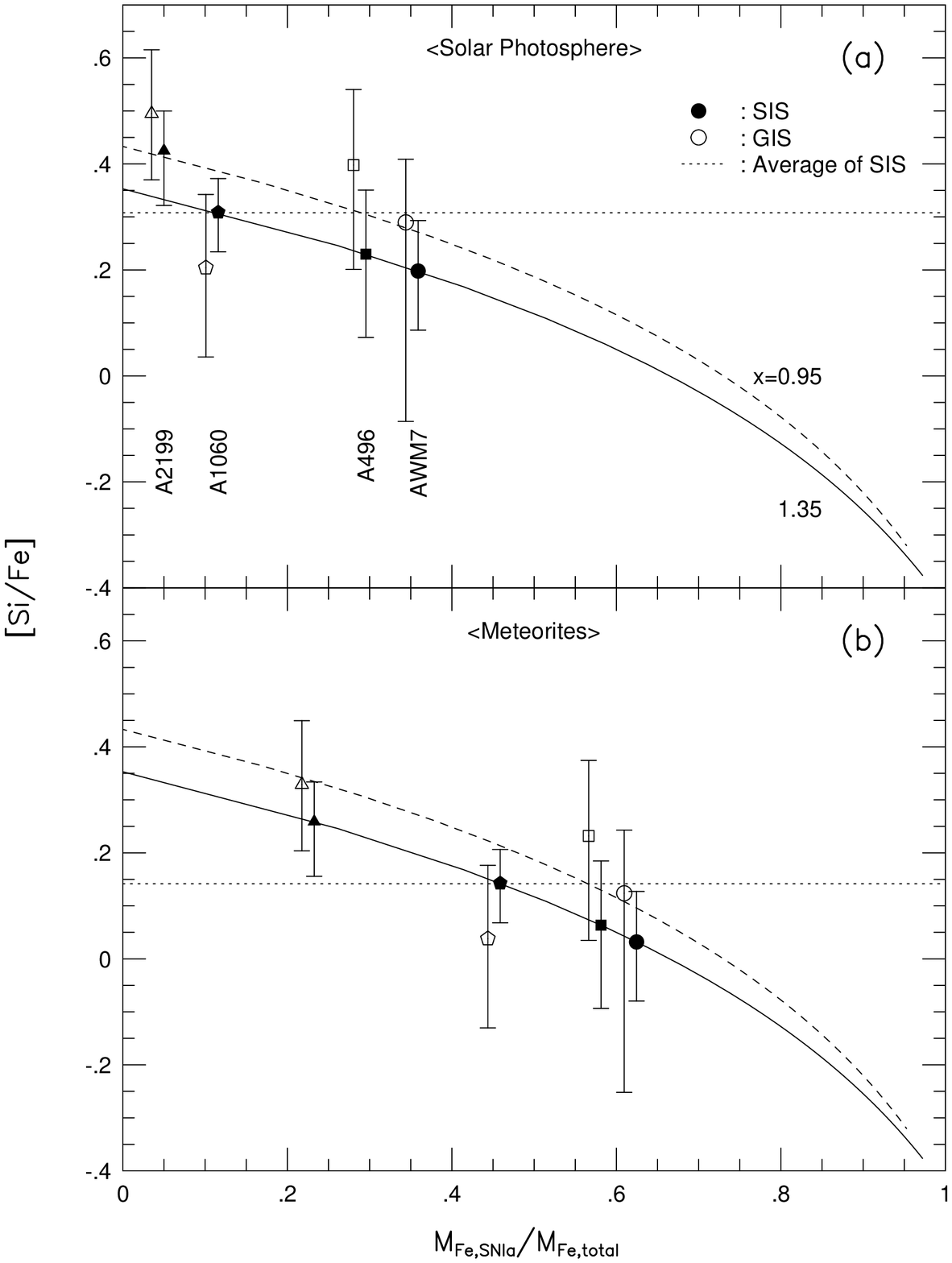}
(a) 
Theoretical ratio of the relative abundance [Si/Fe] as a function of the SNIa
fraction in the iron synthesis. The solid line gives a prediction with the Salpeter
IMF ($x=1.35$), and the dashed line with the flat IMF ($x=0.95$). 
The observed ratios of clusters of galaxies, 
Abell 496 (squares), 1060 (pentagons), 2199 (triangles), and AWM7 (circles)
are taken from Mushotzky et al. (1996). 
The open and closed symbols indicate the SIS and GIS data of each cluster, 
respectively.
The solar photospheric abundance of
iron is assumed for both SIS and GIS data of ASCA. The dotted line gives
an average of the SIS data. 
The original [Si/Fe] values for the four clusters of galaxies 
given by Mushotzky et al. (1996) are superposed
in such a way that the theoretical locus with $x=1.35$ gives a good fit
to each of the SIS data. Since the SIS value of Abell 2199 is much higher than
the theoretical curve, we simply locate it at the left-hand edge of the
figure. The error bars give the 90\% confidence intervals.
(b) Same as figure 1a, but the observational data are corrected by using 
the meteoritic iron abundance.
The error bars give the 90\% confidence intervals.
\end{fv}


\begin{Fv}{2}{41pc}
\includegraphics{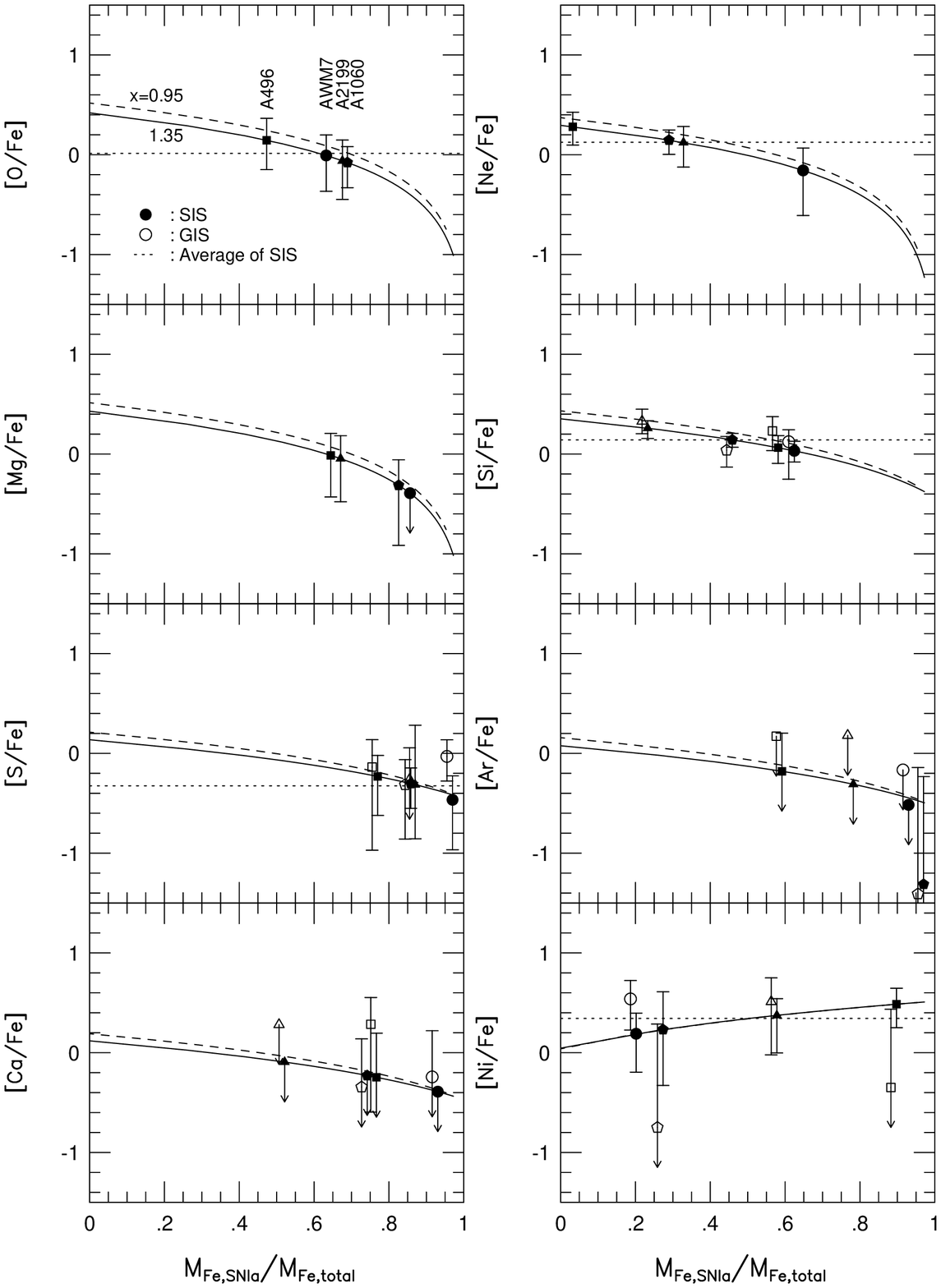}
Same as figure 1b, but for various $\alpha$-elements and nickel. The
observational data were corrected using the meteorite iron abundance.
The error bars give the 90\% confidence intervals.
\end{Fv}


Figures 1a and 1b illustrate the relative-abundance ratio of 
silicate to iron as a function of the fractional SNIa contribution. 
The error bars give 90\% confidence levels.
Silicate is chosen because it is an unique $\alpha$-element for which X-ray 
spectra models can give the abundance most accurately. 
In figure 1a, the original [Si/Fe] values for the four clusters of galaxies 
are superposed
in such a way that the theoretical locus with $x=1.35$ gives a good fit
to each of the SIS data. Since the SIS value of Abell 2199 is much higher than
the theoretical curve, we simply locate it at the left-hand edge of the figure. 
Although figure 1b is the same as figure 1a, the [Si/Fe] 
values are corrected using the meteoritic abundances.
We first note that if [Si/Fe] $\simeq$ 0, as is the case in the Sun, 
$\sim$ 60\% of the iron comes from SNIa (Tsujimoto et al. 1995). 
With the Salpeter IMF, the average of ASCA 
SIS data (dotted line) suggests that only 10\% of the iron comes from 
SNIa if the solar photosphere abundance is assumed, while if the 
meteoritic abundance is adopted, SNIa produces more than 45\% iron 
in the ICM. \ The SNIa contribution further increases by $\sim$ 10\% 
if a flat IMF ($x=0.95$) is assumed.

Figure 2 shows the [Z$_i$/Fe] ratios for O, Ne, Mg, Si, S, Ar, Ca, 
and Ni. \ The error bars give the 90\% confidence levels.
\ Except for Ni, all of the elements are the so-called 
$\alpha$-elements, and the theoretical loci show a similar trend to
that of
the fractional contribution of SNIa. GIS data were not available 
for Ne, Mg, Ar, and Ca. \ The accuracy of the SIS data for Mg, Ar, and Ca are 
very poor (Mushotzky et al. 1996); these elements were therefore not 
considered in this study. \ Although Ni is shown for a comparison, it does 
not provide any information about the SNIa fraction, because both Ni
and Fe are mainly produced in SNIa. \ The [O/Fe] ratios give 
$M_{\rm Fe,SNIa}/M_{\rm Fe,total}=$ 0.45--0.70, similar
to that derived from [Si/Fe]. [S/Fe] gives an even higher value of 
$M_{\rm Fe,SNIa}/M_{\rm Fe,total} \ge 0.75$.
\ Although [Ne/Fe] suggests a lower value, $\sim$ 0.05--0.65,
the observed neon abundances are uncertain, because
the line is affected by the present-day uncertainties of the Fe-L lines 
(Mushotzky et al. 1996; Loewenstein, Mushotzky 1996;
Arimoto et al. 1997). From these results, 
especially from [Si/Fe] and [O/Fe], we conclude that 
an SNIa iron fraction of 50\% or higher
could not be ruled out and might actually be favoured based on the
ASCA spectra.
The implication of this result is significant, and an extensive
theoretical discussion concerning iron production in clusters of
galaxies will be publised elsewhere by us.

\section{Discussions}

It has become a general consensus that iron in the ICM was produced 
in elliptical galaxies and ejected via an SN-driven wind and/or 
outflows (Arimoto, Yoshii 1987; Matteucci, Tornamb\'e 1987; 
Matteucci, Vettolani 1988); still, the relative role of SNIa and 
SNII remains a matter of debate (Matteucci 1992; David et al. 
1991a,b; Ciotti et al. 1991; Arnaud et al. 1992; Renzini et al. 
1993; Elbaz et al. 1995; Matteucci, Gibson 1995; Gibson 1996; 
Loewenstein, Mushotzky 1996). 

To account for the IMLR of clusters
and the observed properties of nearby elliptical galaxies, 
Elbaz et al. (1995) have recently claimed that bimodal star formation in 
elliptical galaxies, originally proposed by Arnaud et al. (1992), is 
responsible for the iron enrichment of the ICM. 
\ In this scenario, only 
high-mass stars are born in the violent star-formation phase during the initial
stage of evolution. \ The SNII products were ejected through a wind, followed
by a more quiescent formation of stars with the normal IMF. 
\ In this way, the SNIa events are greatly suppressed,
and almost all iron in the ICM has an origin in SNII. 
\ As a natural consequence of this scenario, Elbaz et al.
predict 
that the $\alpha$-elements to the iron ratios should be above solar in the ICM: 
[O/Fe]=+0.32, [Si/Fe]=+0.11, and [Mg/Fe]=+0.26.
\ Since the meteoritic abundances of Anders and Grevesse (1989)
are used as the solar abundances in their chemical-evolution model, 
these predictions should be compared directly to those given in table 1.
Obviously, their model fails to reproduce these ASCA data.
Moreover, we note that Elbaz et al. (1995) 
assume the mean iron abundance of the stars 
in ellipticals to be [Fe/H]=$+$0.15, according to Buzzoni et al. (1992). This
value is over-estimated, because the Mg$_2$ indices studied by Buzzoni et al. were
taken from the central parts of ellipticals, while ellipticals
generally 
show a strong gradient in the metallic line indices, including Mg$_2$ (e.g., Davies
et al. 1993; Gonzalez 1993; Carollo et al. 1995). By using the observational
data of the Mg$_2$ gradient for about 50 ellipticals, Arimoto et al. (1997)
have estimated the mean stellar iron abundance for each galaxy, and find that 
luminous giant ellipticals have at most [Fe/H] $\simeq$ 0, and less luminous
ones a much smaller value. Since [Fe/H]=$+$0.15 
is a direct consequence of their model, 
a proper account of the metallicity gradient in ellipticals would require
a lower amount of locked-up iron 
produced by SNII, which in turn implies an additional
source of iron, i.e., SNIa, to quantitatively explain the iron 
mass in the ICM.

A proper chemical model should also accommodate the iron abundance of hot 
X-ray gas in an elliptical-galaxy halo. 
Although ASCA has detected a surprisingly low iron abundance,
less than half solar or even less
(Awaki et al. 1994; Loewenstein et al. 1994; Mushotzky et al. 1994),
the interpretation of the iron-L lines is still not well understood
(Arimoto et al. 1997). For the same reason, a recent claim that
[$\alpha$/Fe] of the hot gas in ellipticals is sub-solar, implying the dominant
contribution from SNIa, needs to be confirmed 
(e.g., Awaki et al. 1994).

Loewenstein and Mushotzky (1996) discussed the IMLR of O, Si, and Fe
in addition to the relative abundance ratios ([O/Fe], [Ne/Fe], 
[Si/Fe], and [S/Fe]) with a help of an ICM enrichment model
in which only the SNII nucleosynthesis was taken into account.
Since the solar photospheric value was assumed 
in the ASCA data analyses, the same {\it high} iron
abundance was adopted in their model.
\ If the ICM is enriched by SNII alone, the relative abundance ratios
studied by Loewenstein and Mushotzky (1996) decrease as a function
of the IMF slope $x$. They therefore found that the abundance ratios
could be explained by SNII alone, provided that the IMF has a
very steep slope of $x \sim 1.5$. 
Since the stellar yields given by the IMF 
with such a steep slope is too low to explain the observed IMLR of
O, Si, and Fe, the authors claimed 
that the ICM must have been enriched by a significant number of 
SNII stars during the initial stage of evolution, 
where only high-mass stars were formed, i.e., bimodal star
formation. However, SNIa were entirely ignored in their model,
and indeed the solution to the problem is quite different
if both SNIa and SNII are properly taken into account.

Following Tsujimoto et al. (1995), we have assumed the mass range of 
SNII progenitors to be $10 \le m/M_\odot \le 50$. \ Stars of 8--10$M_\odot$
produce negligible iron, a crucial element in our discussion,
through O-Ne-Mg core collapse (Hashimoto et al. 1993). However, there
still remain several uncertainties in the present day SNII nucleosynthesis:
1) Nucleosynthesis prescriptions 
for $10\le m/M_\odot < 13$ are not available.
Tsujimoto et al. (1995) assumed that the stellar yields decrease
linearly with mass from $13M_\odot$ to $10M_\odot$, which we also
adopted in this study. Since this might underestimate the iron production from 
stars of this mass range, we repeated our calculation by
assuming the same stellar yields for 
$10 \le m/M_{\odot} < 13$ as those of 13$M_\odot$.
A fit in the $M_{\rm Fe,SNIa}/M_{\rm Fe,total}$ vs [Si/Fe] diagram
was obtained at $M_{\rm Fe,SNIa}/M_{\rm Fe,total} \simeq 0.30$ if $x=1.35$ and 
0.45 if $x=0.95$. Thus, a rather low SNIa contribution is suggested.
However, this is an extreme case, which
gives a considerably lower [O/Fe] ratio than those observed for halo
giants in our Galaxy (Tsujimoto et al. 1995).
2) The light curve of SN 1993J (SNII) suggests that the iron production of 
13--15$M_\odot$ stars should be $\sim 0.1 M_\odot$ 
(Nomoto 1996, private communication), while the values predicted from
a nucleosynthesis study are around $0.15M_\odot$ (Thielemann et al. 1996).
We therefore calculated the SNIa contribution while assuming that 13--15$M_\odot$
stars produce 0.1$M_\odot$ iron, and keeping the stellar yields of 
all other elements the same as that of Thielemann et al. (1996). The resulting 
$M_{\rm Fe,SNIa}/M_{\rm Fe,total}$ is $\simeq 0.60$ if $x=1.35$ and 
0.65 if $x=0.95$. Obviously, SNIa contributes significantly in this case, 
because these stars dominate the SNII nucleosynthesis, and even a slight
decrease in the stellar iron yield crucially reduces the SNII contribution
to the ICM enrichment.
3) Although Tsujimoto et al. (1995) found $50M_\odot$ as the best 
fit of their chemical-evolution model to the solar neighbourhood,
the precise value of the lower mass limit for black hole formation
is still quite uncertain. We calculated the SNIa fraction by
assuming $m_{\rm u} = 70 M_\odot$ instead of $50M_\odot$. The resulting
$M_{\rm Fe,SNIa}/M_{\rm Fe,total}$ is $\simeq 0.60$ if $x=1.35$ and 
0.70 if $x=0.95$. It is clear from these calculations that
our conclusions would not be seriously influenced by the uncertainties 
involved in the present-day SNII nucleosynthesis prescriptions.

\section{Conclusions}

A long-outstanding disagreement concerning the solar 
photospheric iron abundance has recently 
converged to the so-called {\it low} value (Bi\'emont et al. 1991; Holweger 
et al. 1991), which agrees well with the meteoritic one (Anders,
Grevesse 1989). Mushotzky et al. (1996) recently determined 
the abundances of O, Ne, Mg, Si, S, Ar, Ca, Ni, and Fe in a sample of 
four rich 
clusters of galaxies. They found that the abundance pattern of each cluster is
very similar, and 
suggested that the ratio of the relative abundance of the elements is
consistent with the origin of most of 
the heavy elements in SNII. \ However, their
results were derived based on the use of the so-called {\it high} solar iron
abundance taken from the old solar photospheric data 
(Anders, Grevesse 1989). 
\ We have corrected Mushotzky et al.'s abundances by using the meteoritic
values, and have estimated the relative role of SNIa and SNII in the ICM with
the help of the latest nucleosynthesis prescriptions for both SNIa and SNII. 
We find that 
an SNIa iron fraction of 50\% or higher
could not be ruled out, and might actually be favoured by the
ASCA spectra.
This result is rather contrary to what Mushotzky et al. suggest, and 
does not support the so-called bimodal IMF scenario proposed
by Arnaud et al. (1992) and discussed extensively by Elbaz et al. (1995).

\vspace{1pc} \par
We thank to an annonymous referee for her/his constructive suggestions,
which helped us considerablly to improve the original version
of the paper.
We are grateful to K. Nomoto for his continuous encouragement throughout
the course of this investigation. We also thank to S. Wanajo for useful
discussions. YI acknowledges financial support from 
the Japan Society for Promotion of Science. 
NA thanks to the Institut f\"ur Theoretische Astrophysik, Universit\"at
Heidelberg and to Sonderforschungsbereich (SFB) 328 
and the Physics of Department, University of Durham
for financial support while the major parts of this work were proceeding.
This work was financially supported in part by a Grant-in-Aid for the
Scientific Research (No.06640349 and No.07222206) by the Japanese
Ministry of Education, Science, Sports and Culture.

\vspace{1pc} \par
\section*{References}

\re{Allen C.W. 1973, Astrophysical Quantities (Athlone, London)}
\re{Anders E., Ebihara M. 1982, Geochim. Cosmochim. Acta 46, 2363} 
\re{Anders E., Grevesse N. 1989, Geochim. Cosmochim. Acta 53, 197}
\re{Arimoto N., Matsushita K., Ishimaru Y., Ohashi T., Renzini A.
   1997, ApJ in press}
\re{Arimoto N., Yoshii Y. 1987, A\&A 173, 23}
\re{Arnaud M., Rothenflug R., Boulade O., Vigroux L., Vangioni-Flam E. 
   1992, A\&A 254, 49}
\re{Awaki H., Mushotzky R.F., Tsuru T., Fabian A.C., Fukazawa Y.,
    Lowenstein M., Makishima K., Matsumoto H., et al. 1994, PASJ 46, L65} 
\re{Bi\'emont E., Baudoux M., Kurucz R.L., Ansbacher W., Pinnington E.H.
    1991, A\&A 249, 539}
\re{Blackwell D.E., Booth A.J., Petford A.D. 1984, A\&A 132, 236}
\re{Buzzoni A., Gariboldi G., Mantegazza L. 1992, AJ 103, 1814}
\re{Cameron A.G.W. 1970, Space Sci. Rev. 15, 121}
\re{Cameron A.G.W. 1982, 
    in Essays in Nuclear Astrophysics, 
    ed C.A. Barnes, D.D. Clayton, D.N. Schramm
    (Cambridge Univ. Press, Cambridge) p23}
\re{Carollo C.M., Danziger I.J., Buson L. 1995, MNRAS 265, 553}
\re{Ciotti L., D'Ercole A., Pellegrini S., Renzini A. 1991, ApJ 376,
   380} 
\re{David L.P., Forman W., Jones C. 1991a, ApJ 369, 121}
\re{David L.P., Forman W., Jones C. 1991b, ApJ 380, 39}
\re{Davies R.L., Sadler E., Peletier R. 1993, MNRAS 262, 650}
\re{Edge A., Stewart G. 1991, MNRAS 252, 414}
\re{Edvardsson B., Andersen J., Gustafsson B., Lambert D.L., 
    Nissen P.E., Tomkin J.
    1993, A\&A 275, 101} 
\re{Elbaz D., Arnaud M., Vangioni-Flam E. 1995, A\&A 303, 345}
\re{Gibson B.K. 1996, MNRAS 278, 829}
\re{Gonzalez J. 1993, PhD Thesis, University of California at Santa Cruz}
\re{Greggio L., Renzini A. 1983, A\&A 118, 217}
\re{Grevesse N. 1984a, 
    Abundances of the elements in the Sun,
    in Frontiers of Astronomy and Astrophysics,
    ed R. Pallavicini,
    (Ital. Astron. Soc. Florence, Italy) p71}
\re{Grevesse N. 1984b, Physica Scripta T8, 49}
\re{Hashimoto M., Iwamoto K., Nomoto K. 1993, ApJ 414, L105}
\re{Hatsukade I. 1989, Ph.D. Thesis, Osaka University} 
\re{Holweger H., Bard A., Kock A., Kock M. 1991, A\&A 249., 545}
\re{Holweger H., Heise C., Kock M. 1990, A\&A 232, 510}
\re{Ikebe Y., Ohashi T., Makishima K., Tsuru T., Fabbiano G.,
    Kim D.-W., Trinchieri G., Hatsukade I., et al.
    1992, ApJ 384, L5}
\re{King J.R. 1993, AJ 106, 1206}
\re{Larson R.B. 1974, MNRAS 169, 229}
\re{Loewenstein M., Mushotzky F. 1996, ApJ 466, 695}
\re{Loewenstein M., Mushotzky F., Tamura T., Ikebe Y.,
    Makishima K., Matsushita K., Awaki H., Serlemitsos P.J. 1994, ApJL 436, L75}
\re{Matteucci F. 1992, ApJ 397, 32}
\re{Matteucci F., Fran\c cois P. 1992, A\&A 262, L1}
\re{Matteucci F., Gibson B.K. 1995, A\&A 304, 11.}
\re{Matteucci F., Tornamb\`e A. 1987, A\&A 185, 51}
\re{Matteucci F., Vettolani G. 1988, A\&A 202, 21}
\re{Mihara K., Takahara F. 1994, PASJ 46, 447}
\re{Mushotzky R., Loewenstein M., Arnaud K.A., Tamura T., 
    Fukazawa Y., Matsushita K., Kikuchi K., Hatsukade I.
    1996, ApJ 466, 686}
\re{Mushotzky R., Loewenstein M., Awaki H., Makishima K., 
    Matsushita K., Matsumoto H. 1994, ApJL 436, L79}
\re{Nissen P.E., Gustafsson B., Edvardsson B., Gilmore G. 
    1994, A\&A 285, 440}
\re{Nomoto K., Thielemann F.-K., Yokoi K. 1984, ApJ 286, 644}
\re{Norris J., Peterson R.C., Beers T.C. 1993, ApJ 415, 797}
\re{Pauls U., Grevesse N., Huber M.C.E. 1990, A\&A 231, 536}
\re{Renzini A., Ciotti L., D'Ercole A., Pellegrini S. 1993, ApJ 419, 52}
\re{Rothenflug R., Arnaud M. 1985, A\&A 144, 431}
\re{Thielemann F.-K., Nomoto K., Hashimoto M. 1993, 
    in Origin and Evolution of the Elements,
    ed N. Prantzos, E. Vangoni-Flam
    (Cambridge University Press, Cambridge) p297}
\re{Thielemann F.-K., Nomoto K., Hashimoto M. 1996, ApJ 460, 408}
\re{Timmes F.X., Woosley S.E., Weaver T.A. 1995, ApJS 98, 617}
\re{Tinsley B.M. 1980, Fundam. Cosmic Phys. 5, 287}
\re{Tsujimoto T., Nomoto K., Hashimoto M., Thielemann F.-K. 1994, 
    in Evolution of the Universeand its Observational Quest,
    ed K. Sato (Universal Academy Press, Tokyo) p553)}
\re{Tsujimoto T., Nomoto K., Yoshii Y., Hashimoto M., Yanagida S.,
    Thielemann F.-K. 1995, MNRAS 277, 945}
\re{Tsuru T. 1993, Ph.D. Thesis, The University of Tokyo, ISAS Research Note 528}
\re{Wheeler J.C., Sneden C., Truran J.W. 1989, ARA\&A 27, 279}

\vfill\eject
\end{document}